\documentclass[useAMS,usenatbib]{mn2e}

\usepackage{amssymb}
\usepackage{amsfonts}
\usepackage{natbib}
\usepackage{epsfig}
\usepackage{mncite}

\def\gs{\mathrel{\raise0.35ex\hbox{$\scriptstyle >$}\kern-0.6em 
\lower0.40ex\hbox{$\scriptstyle \sim$}}}
\def\ls{\mathrel{\raise0.35ex\hbox{$\scriptstyle <$}\kern-0.6em 
\lower0.40ex\hbox{$\scriptstyle \sim$}}}
\newcommand{\msun}{M_\odot}
\newcommand{\kms}{\rm kms^{-1}}
 \newcommand{\be}{\begin{equation}}
\newcommand{\ee}{\end{equation}}
\newcommand{\bea}{\begin{eqnarray}}
\newcommand{\eea}{\end{eqnarray}}

\begin{document}

\title[Is there an upper limit to black hole masses?]{Is there an
  upper limit to black hole masses?}

\author[Natarajan \& Treister] {Priyamvada 
Natarajan$^{1,2,3}$ and Ezequiel Treister$^4$\\
$^1$ Department of Astronomy, Yale University, P. O. Box 208101, 
New Haven, CT 06511-208101, USA \\
$^2$ Department of Physics, Yale University, P. O. Box 208120, 
New Haven, CT 06520-208120, USA\\
$^3$ Radcliffe Institute for Advanced Study, Harvard University, 
10 Garden Street, Cambridge, MA 02138, USA\\
$^4$ European Southern Observatory, Casilla 19001, Santiago 19, Chile}

\maketitle

\begin{abstract}
  We make a case for the existence for ultra-massive black holes
  (UMBHs) in the Universe, but argue that there exists a likely upper
  limit to black hole masses of the order of $M \sim 10^{10}
  \msun$. We show that there are three strong lines of argument that
  predicate the existence of UMBHs: (i) expected as a natural
  extension of the observed black hole mass bulge luminosity relation,
  when extrapolated to the bulge luminosities of bright central
  galaxies in clusters; (ii) new predictions for the mass function of
  seed black holes at high redshifts predict that growth via accretion
  or merger-induced accretion inevitably leads to the existence of
  rare UMBHs at late times; (iii) the local mass function of black
  holes computed from the observed X-ray luminosity functions of
  active galactic nuclei predict the existence of a high mass tail in
  the black hole mass function at $z = 0$. Consistency between the
  optical and X-ray census of the local black hole mass function
  requires an upper limit to black hole masses. This consistent
  picture also predicts that the slope of the $M_{\rm bh}$-$\sigma$
  relation will evolve with redshift at the high mass end. Models of
  self-regulation that explain the co-evolution of the stellar
  component and nuclear black holes naturally provide such an upper
  limit. The combination of multi-wavelength constraints predicts the
  existence of UMBHs and simultaneously provides an upper limit to
  their masses. The typical hosts for these local UMBHs are likely the
  bright, central cluster galaxies in the nearby Universe.
  \end{abstract}


\section{Introduction}

Observations of black hole demographics locally is increasingly
providing a strong constraint on models that explain the assembly and
growth of black holes in the Universe. The existence of a tight
relation between the velocity dispersion of bulges and the mass of the
central black hole has been reported by several authors (Merritt \&
Ferrarese 2001; Tremaine et al. 2002; Gebhardt et al. 2003). This
correlation is tighter than that between the luminosity of the bulge
and the mass of the central black hole (Magorrian et al. 1998).  The
physical processes that set up this correlation are not fully
understood at the present time, although there are several proposed
explanations that involve the regulation of star formation with black
hole growth and assembly in galactic nuclei (Haehnelt, Natarajan \&
Rees 1998; Natarajan \& Sigurdsson 1998; Silk \& Rees 1999; Murray,
Quataert and Thompson 2004; King 2005).

Recent work by several authors has suggested that UMBHs\footnote{Black
  holes with masses in excess of $5 \times 10^9\,M_{\odot}$ are
  hereafter referred to as UMBHs.} ought to exist: Bernardi et
al. (2006) show that the high velocity dispersion tail of the velocity
distribution function of early-type galaxies constructed from the
Sloan Digital Sky Survey (SDSS) had been under-estimated in earlier
work suggestive of a corresponding high mass tail for the central
black hole masses hosted in these nuclei. As first argued by Lauer 
et al. (2007a) and subequently by Bernardi et al. (2007) and
Tundo et al. (2007), even when the scatter in the observed
$M_{\rm bh} - \sigma$ correlation is taken into account it predicts
fewer massive black holes compared to the $M_{\rm bh} - L_{\rm bulge}$
relation. While Bernardi et al. (2007) argue that this is due to the fact that the $\sigma -
L_{\rm bulge}$ relation in currently available samples is inconsistent
with the SDSS sample from which the distributions of $L_{\rm bulge}$
or $\sigma$ are based. From an early-type galaxy sample observed by HST, 
Lauer et al. (2007b) argue that the relation between $M_{\rm bh} - L_{\rm bulge}$
 is likely the preferred one for BCGs (Brightest
Cluster Galaxies) consistent with the harboring of UMBHs as evidenced by their large 
core sizes. The fact that the high mass end of the observed
local black hole mass function is likely biased is a proposal that
derives from optical data. Deriving the mass functions of accreting
black holes from optical quasars in the Sloan Digital Sky Survey Data
Release 3 (SDSS DR3), Vestergaard et al. (2008) also find evidence for
UMBHs in the redshift range $0.3 \leq z \leq 5$.

In this paper, we show that UMBHs exist using X-ray and bolometric AGN
luminosity functions and for consistency with local observations of
the BH mass density, an upper limit to their masses is required. To
probe the high mass end of the BH mass function, in earlier works the
AGN luminosity functions were simply extrapolated. This turns out to
be inconsistent with local estimates of the BH mass function. Here we
focus on the high mass end of the predicted local black hole mass
function, i.e.  extrapolation of the $M_{\rm bh} - \sigma$ relation to
higher velocity dispersions and demonstrate that a self-limiting
cut-off in the masses to which BHs grow at every epoch reconciles the
X-ray and optical views.

The outline of this paper is as follows: in Section 2, we briefly
summarise the current observational census of black holes at high and
low redshift including constraints from X-ray AGN. The pathways to
grow UMBHs are described in Section 3. Derivation of the local black
hole mass function from the X-ray luminosity functions of AGN is
presented in Section 4. The arguement for the existence of an upper
limit to black hole masses from various lines of evidence is presented
in Section 5; the prospects for detection of this population is
presented in Section 6 followed by conclusions and discussion. We
adopt a cosmological model that is spatially flat with $\Omega_{\rm
matter} = 0.3$; $H_0 = 70\,{\rm km~s^{-1}}/{\rm Mpc}$.

\section{Status of current census of black holes at high and low redshift}

\begin{figure}
\begin{center}
\includegraphics[height=9cm,width=9cm]{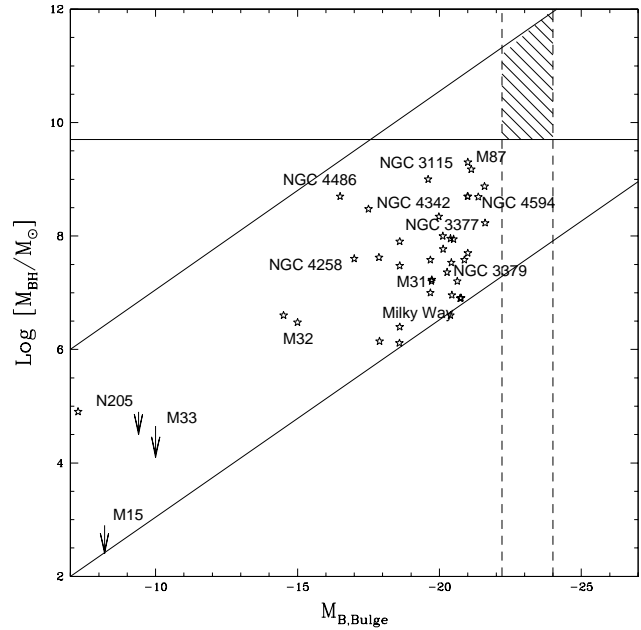}
\caption{Relation between the inferred black hole mass vs. the host
bulge luminosity; data taken from Magorrian et al. (1998); Ho (1998)
and Gebhardt et al. (2002). The vertical dashed lines indicate the
typical luminosity of cD galaxies and the hatched region is the
parameter space for finding UMBHs.}
\end{center}
\end{figure}

The demography of local galaxies suggests that every galaxy hosts
a quiescent supermassive black hole (SMBH) at the present time and the
properties of the black hole are correlated with those of the host. In
particular, observational evidence points to the existence of a
strong correlation between the mass of the central black hole and the
velocity dispersion of the host spheroid (Tremaine et al. 2002;
Merritt \& Ferrarese 2001, Gebhardt et al. 2002) in nearby
galaxies. This correlation strongly suggests coeval growth of the
black hole and the stellar component via likely regulation of the gas
supply in galactic nuclei (Silk \& Rees 1999; Kauffmann \& Haehnelt
2000; Cattaneo 2001; Bromley, Somerville \& Fabian 2004; King 2003;
Murray, Quataert \& Thompson 2005; Sazonov et al. 2005; Begelman \&
Nath 2005; Alexander et al. 2005).

Black hole growth is primarily powered by gas accretion (Lynden-Bell
1969) and accreting black holes that are optically bright are detected
as quasars. The build-up of SMBHs is likely to have commenced at
extremely high redshifts. Indeed, optically bright quasars have now
been detected at $z > 6$ (e.g., Fan et al.\ 2001a, 2003) in the SDSS.
There are also indications that high redshift quasar hosts are strong
sources of dust emission (Omont et al.\ 2001; Cox et al.\ 2002;
Carilli et al.\ 2002; Walter et al.\ 2003; Reuland et al. 2004),
suggesting that quasars were common in massive galaxies at a time when
galaxies were undergoing copious star formation. The growth spurts of
SMBHs are also detected in the X-ray waveband. The summed emission
from these AGN generates the cosmic X-ray Background (XRB), and its
spectrum suggests that most black-hole growth is optically obscured
(Fabian 1999; di Matteo et al. 1999; Mushotzky et al. 2000; Hasinger
et al. 2001; Barger et al. 2003; Barger et al. 2005; Worsley et
al. 2005). There are clear examples of obscured black-hole growth in
the form of `Type-2' quasars, and the detected numbers are in
agreement with some recent XRB models (Treister \& Urry 2005; Gilli et
al. 2007) and have the expected luminosity dependence of the obscured
fraction.  Additionally, there is tantalizing recent evidence from
infra-red (IR) studies that dust-obscured accretion is ubiquitous
(Martinez-Sansigre et al. 2005, 2007). At present it is unknown what
fraction of the total mass growth occurs in such an optically dim
phase as a function of redshift.

The build-up of BH mass in the Universe has been traced using optical
quasar activity. The current phenomenological approach to
understanding the assembly of SMBHs involves optical data from both
high and low redshifts. These data are used to construct a consistent
picture that fits within the larger framework of the growth and
evolution of structure in the Universe (Haehnelt, Natarajan \& Rees
1998; Haiman \& Loeb 1998; Kauffmann \& Haehnelt 2000; 2002; Wyithe \&
Loeb 2002; Volonteri et al. 2003; Di Matteo et al. 2003; Steed \&
Weinberg 2004).

Black hole accretion histories derived from the quasar luminosity
function (e.g. Soltan 1982; Haehnelt, Natarajan \& Rees 1998; Salucci
et al. 1999; Yu \& Tremaine 2002; Marconi et al. 2004; Shankar et
al. 2004; Merloni et al. 2004), synthesis models of the XRB
(e.g. Comastri et al. 1995; Gilli et al. 1999; Elvis et al. 2002; Ueda
et al. 2003; Barger et al. 2005; Treister \& Urry 2005; Gilli et
al. 2007), and observations of accretion rates in quasars at different
redshifts (Vestergaard 2004; McLure \& Dunlop 2004) and composite
models (Hopkins et al. 2005b; 2006a; 2006b) suggest that supermassive
black holes spend most of their lives in a low efficiency, low
accretion rate state. In fact, only a small fraction of the SMBHs
lifetime is spent in the optically bright quasar phase, although the
bulk of the mass growth occurs during these epochs. In this paper, we
examine the consequences of such an accretion history for the high
mass end of the local black hole mass function.
 
Surveys at X-ray energies allow us to obtain a more complete view of
the AGN population, as they cover a broader range in luminosity and
are simultaneously less affected by baises due to obscuration. While
optical surveys of quasars, like the SDSS or 2dF, are used to obtain a
large sample of unobscured and high-luminosity sources, it is with
X-ray surveys that the obscured low-luminosity population can be well
traced. In particular, surveys at hard X-ray energies, 2--10~keV, are
almost free of selection effects up to columns of $N_H \sim
10^{23}\,{\rm cm}^{-2}$. In the work of Ueda et al. (2003) the AGN
X-ray luminosity function is computed based on a sample of $\sim$250
sources observed with various X-ray satellites. One of the important
conclusions of this paper is the confirmation of a
luminosity-dependent density evolution, in the sense that lower
luminosity sources peak at lower redshifts, $z\,<\,1$, while only the
high luminosity sources are significantly more abundant at
$z\,\sim\,2$, as observed in optical quasar surveys (e.g., Boyle et
al. 2000). Additionally, using this X-ray luminosity function and
evolution it was possible for Ueda et al. (2003) to convincingly
account for the observed properties of the extragalactic XRB.

Extending the argument presented by Soltan (1982) to the X-ray
wave-band, AGN activity can be used to trace the history of mass
accretion onto supermassive black holes (Fabian \& Iwasawa
1999). Marconi et al. (2004) and Shankar et al. (2004) used the
luminosity function of Ueda et al. (2003) to calculate the spatial
density of supermassive black holes inferred from AGN activity and
compared that with observations. These authors reported in general a
good agreement between observations and the density inferred from AGN
relics, suggesting that there is little or no room for further
obscured accretion, once Compton thick AGN are properly accounted
for. A similar conclusion was also obtained by Barger et al. (2005)
from an independently determination of the luminosity function, thus
confirming this result.

\section{Pathways for growing UMBHs}

Below we discuss plausible scenarios for forming these UMBHs at low
redshift. There are 2 feasible channels for doing so: (i) expect
extremely rare UMBHs to form from the merging of black holes due to
the merging of galaxies via the picture suggested by Volonteri et
al. (2003); (ii) form from accretion onto high redshift `seeds' with
perhaps a brief period of Super-Eddington accretion, the descendants
of the SMBHs that power the most luminous quasars at $z = 6$ as
proposed recently by Volonteri \& Rees (2005); Begelman, Volonteri \&
Rees (2006); Lodato \& Natarajan (2007) and Volonteri, Lodato \&
Natarajan (2007). We discuss these two possible channels for growing
UMBHs in more detail below.

\subsection{Merging history of black holes}

Following the merging DM hierarchy of halos starting with seed BHs at
$z = 20$, populating the $3.5 - 4 \sigma$ peaks, Volonteri et
al. (2003) are able to reproduce the mass function of local BHs as
well as the abundance of the rare $10^9 \msun$ BHs that power the $z =
6$ SDSS quasars. Proceeding to rarer peaks say, $6\sigma$ at $z = 20$
in this scheme yields the rarer $10^{10} \msun$ local UMBHs. And in
fact, the formation of a very small number density of UMBHs at $z = 0$
is inevitable in the standard hierarchical merging $\Lambda$CDM
paradigm. A massive DM halo with mass, $M = 10^{13} \msun$ at $z = 0$
which is the likely host to an UMBH, is likely to have experienced
about 100 mergers between $z = 6$ and $z = 0$, starting with $10^9
\msun$ at $z = 6$.

Recently a numerical calculation of the merger scenario mentioned
above has been performed in simulations by Yoo et al. (2007).
Focusing on the merger history of high mass cluster-scale halos ($M
\sim 10^{15} \,M_{\odot}$). They find that in ten realizations of
halos on this mass scale, starting with the highest initial BH masses
at $z = 2$ of $\sim$ few times $10^9\,M_{\odot}$, 4 clusters contain
UMBHs at $z = 0$.  Therefore, rare UMBHs are expected in the local
Universe. Yoo et al. (2007) argue that black hole mergers can
significantly augment the high end tail of the local BH mass function.

Similarly, using a model for quasar activity based on mergers of
gas-rich galaxies, Hopkins et al. (2006a) showed that they could
explain the observed local BH mass at low to intermediate BH masses
(10$^6$-10$^9$$M_{\odot}$). However, at higher BH masses, their
calculations overpredict the observed values even considering a
possible change in the Eddington fraction at higher masses.

\subsection{Growth from massive high redshift seeds}

Conventional models of black hole formation and growth start with
initial conditions at high redshift with seed BHs that are remnants of
the first generation of stars in the Universe. Propagating these seeds
via merger accompanied accretion events leading to mass growth for the
BHs (Volonteri, Haardt \& Madau 2003) it has been argued that in order
to explain the masses of BHs powering the bright $z \sim 6$ quasars by
the SDSS survey (Fan et al. 2004; 2006) that either a brief period of
Super-Eddington accretion (Volonteri \& Rees 2005) or more massive
seeds are needed (Begelman, Volonteri \& Rees 2006; Lodato \&
Natarajan 2006; Lodato \& Natarajan 2007). Massive seeds can alleviate
the problem of assembling $\sim 10^9\,M_{\odot}$ BHs by $z = 6$ which
is roughly 1 Gyr after the Big Bang in the concordance $\Lambda$CDM
model. The local relics of such super-grown black holes are expected
to result in UMBHs. We note here that following the evolution of the
massive black holes that power the $z =6$ quasars, in a cosmological
simulation, Di Matteo et al. (2008) find that these do not necessarily
remain the most massive black holes at subsequent times. Therefore,
while UMBHs might not be direct descendants of the SMBHs that power
the $z=6$ quasars, there is ample room for UMBHs to form and grow.

Below, we briefly present scenarios that provide the massive BH seeds
in the first place that will eventually result in a small population
of UMBHs by $z = 0$. These physically plausible mechanisms are
critical to our prediction of UMBHs at low redshift. Two models have
been proposed, one that involves starting from the remnants of
Population III stars with brief episodes of accretion onto them
exceeding the Eddington rate to bump up their masses (Volonteri \&
Rees 2006) and the other that explains direct formation of massive BH
seeds prior to the formation of the first stars (Lodato \& Natarajan
2006; 2007).

Volonteri \& Rees (2005) have proposed a scenario to explain the high
BH masses $\sim 10^9 \msun$ needed to power the luminous quasars
detected $z = 6$ in the SDSS. This is accomplished they argue by
populating the $4\sigma$ peaks in the dark matter density field at $z
\sim 24$ with seed BHs which arise from the remnants of Population III
stars in the mass ranges $20 \msun < {\rm M}_{\rm bh} < 70 \msun$ and
$130 \msun < {\rm M}_{\rm bh} < 600 \msun$. These remnant BHs then
undergo an episode of super-Eddington accretion from $6 <z < 10$. They
argue that in these high redshift, metal-free dark matter halos $T >
10^4 K$ gas can cool in the absence of $H_2$ via atomic hydrogen lines
to about $8000 K$. As shown by Oh \& Haiman (2002) the gas at this
temperature settles into a rotationally supported `fat' disk at the
center of the halo under the assumption that the DM and the baryons
have the same specific angular momentum. Further, these disks are
stable to fragmentation and therefore do not form stars and
exclusively fuel the BH instead. The accretion is via stable
super-critical accretion at rates well in excess of the Eddington rate
due to the formation of a thin, inner feeding disk. The accretion
radius is comparable to the radiation trapping radius which implies
that all the gas is likely to end up in the BH. Any further cooling
down to temperatures of $10 K < T < 200 K$ for instance, halts the
accretion, causes fragmentation of the disk which occurs when these
regions of the Universe have been enriched by metals.  This process
enables the comfortable formation of $10^9 \msun$ BHs by $z = 6$ or so
to explain the observed SDSS quasars. In a $\Lambda$CDM Universe, the
time available from $z = 6$ to $z = 0$ is $\sim 12.7$ Gyr. To grow by
an order of magnitude during this epoch requires an accretion rate of
$< 1 \msun {\rm yr}^{-1}$ which is well below the Eddington rate;
however, it requires a gas rich environment.

In recent work, Lodato \& Natarajan (2007) have shown that an
ab-initio prediction for the mass function of seed black holes at high
redshift can be obtained in the context of the standard $\Lambda$CDM
paradigm for structure formation combined with careful modeling of the
formation, evolution and stability of pre-galactic disks. They show
that in dark matter halos at high redshifts $z \sim 15$, where zero
metallicity pre-galactic disks assemble (prior to the formation of the
first stars), gravitational instabilities in these disks transfer
angular momentum out and mass inwards efficiently.  Note that the only
coolants available to the gas at this epoch are either atomic or
molecular hydrogen.  Taking into account the stability of these disks,
in particular the possibility of fragmentation, the distribution of
accumulated central masses in these halos can be computed. The central
mass concentrations are expected to form seed black holes. The
application of stability criteria to these disks leads to distinct
regimes demarcated by the value of the $T_{\rm vir}/T_{\rm gas}$ where
$T_{\rm gas}$ is the temperature of the gas and $T_{\rm vir}$ is the
virial temperature of the halo. The three regimes and consequences are
as follows: (i) when $T_{\rm vir}/T_{\rm gas}\,> 3$ the disk fragments
and forms stars instead of a central mass concentration; (ii)
$2\,<\,T_{\rm vir}/T_{\rm gas}\,<\,3$, when both central mass
concentrations and stars form ; (iii)$T_{\rm vir}/T_{\rm gas}< 2$,
when only central mass concentrations form and the disks are stable
against fragmentation. Using the predicted mass function of seed black
holes at $z \sim 15$, and propagating their growth in a merger driven
accretion scenario we find that the masses of black holes powering the
$z=6$ optical quasars can be comfortably accommodated and consequently
a small fraction of UMBHs is predicted at $z = 0$.  Evolving and
growing these seeds to $z = 0$, the abundance of UMBHs can be
estimated (Volonteri, Lodato \& Natarajan 2008).

\section{The local black hole mass function derived from X-ray luminosity
functions of AGN}

The new evidence that we present in this work for the existence of a
rare population of UMBHs stems from using X-ray luminosity functions
of AGN and the implied accretion history of black holes. Hard X-rays
have the advantage of tracing both obscured and unobscured AGN, as the
effects of obscuration are less important at these energies. In
particular, we use the hard X-ray luminosity function and
luminosity-dependent density evolution presented by Ueda et al. (2003)
defined from $z=0$ out to $z=3$. We further assume that these AGN are
powered by BHs accreting at the Eddington limit. In order to calculate
bolometric luminosities starting from the hard X-ray luminosity the
bolometric corrections derived from the AGN spectral energy
distribution library presented by Treister et al. (2006) are
used. These are based mainly on observations of local AGN and quasars
and depend only on the intrinsic X-ray luminosity of the source, as
they are based on the X-ray to optical ratios reported by Steffen et
al. (2006). To account for the contribution of Compton-thick AGN to
the black hole mass density missed in X-ray luminosity functions, we
use the column density distribution of Treister \& Urry (2005) with
the relative number of Compton-thick AGN adapted to match the spatial
density of these sources observed by INTEGRAL, obtained from the AGN
catalog of Beckmann et al. (2006). In order to account for sources
with column densities $N_H$=10$^{25}$-10$^{26}$~cm$^{-2}$ which do not
contribute much to the X-ray background, but can make a significant
contribution to the BH mass density (e.g., Marconi et al. 2004), we
multiply the BH mass density due to Compton-thick AGN by a factor of
2, i.e., we assume that they exist in the same numbers as in the
$N_H$=10$^{24}$-10$^{25}$~cm$^{-2}$ range, in agreement with the
assumption of Marconi et al. (2004) and consistent with the $N_H$
distribution derived from a sample of nearby AGN by Risaliti et
al. (1999). Under this assumption, the contribution of sources with
$N_H$$>$10$^{25}$~cm$^{-2}$ to the total population of SMBHs is
$\sim$7\%.

We then convert these X-ray LF's to an equivalent BH mass function,
and evolve these mass functions by assuming that accretion continues
at the Eddington rate down to $z=0$. The results of this procedure are
shown in Fig.~2 for three different values of the accretion efficiency
$\epsilon$. Note that we do not consider models in which the
efficiency parameter varies with redshift or BH mass since such models
merely add more unconstrained parameters. As can be seen clearly in
Fig.~2, these simple models do not reproduce the observed local black
hole mass function at the high mass end. The functional form adopted for
the X-ray luminosity function is a double power-law as proposed by
Ueda et al. (2003):
\begin{equation}
\frac{{\rm d} \Phi (L_{\rm X}, z=0)}{{\rm d Log} L_{\rm X}} 
= A [(L_{\rm X}/L_{*})^{\gamma 1} + (L_{\rm X}/L_{*})^{\gamma 2}]^{-1}.
\end{equation}
And the evolution is best described by the luminosity dependent
density evolution model (LDDE model), where the cut-off redshift 
$z_{\rm c}$ is expressed by a power law of $L_{\rm X}$ , consistent with 
observational constraints (see Ueda et al. 2003 for more details):
\begin{equation}
\frac{{\rm d} \Phi (L_{\rm X}, z)}{{\rm d Log} L_{\rm X}} 
= \frac{{\rm d} \Phi (L_{\rm X}, 0)}{{\rm d Log} L_{\rm X}} e(z, L_{\rm X})
\end{equation}
where
\begin{eqnarray}
e(z, L_{\rm X}) = (1+z)^{p1}\,\,\,\,(z<z_{\rm c}(L_{\rm X}))
\end{eqnarray}
\begin{eqnarray}
e(z_{\rm c})[(1+z)/(1+z_{\rm c}(L_{\rm X}))]^{p2}\,\,\,\,\,(z \geq z_{\rm c}(L_{\rm X})).
\end{eqnarray}

This simple and conservative analysis predicts a population of UMBHs
with a local abundance of $\sim$3$\times$10$^{-6}\,{\rm Mpc}^{-3}$!
This is fairly robust as this population is predicted for a large
range of efficiencies. These LF's shown in Fig.~2 also simultaneously
account for the cosmic XRB, as shown by several authors (for instance
see Treister \& Urry (2005) and Gilli et al. (2007) and references
therein), suggesting that the X-ray view presents a fairly complete
picture of the accretion and growth of BHs. Note that our estimates of
the black hole mass function are in general agreement with those of
Marconi et al. (2004) [for a direct comparison see their Fig.~ 2,
right-hand panel], the very slight difference arises due to an
alternate choice of bolometric correction factors and our prescription
for including Compton thick AGN. Estimates by other authors are also
in agreement with our treatment here out to masses of a few times
$10^8\,M_{\odot}$. For BH masses $<\,10^9\,\msun$, there appears to be
consistency between the optical and X-ray views of black hole growth. However for 
$M_{\rm bh}\,>\,10^9\,\msun$, all models that assume Eddington
accretion with varying efficiencies systematically over-estimate the
local abundance of high mass black holes.

\begin{figure}
\begin{center}
\includegraphics[height=9cm,width=9cm]{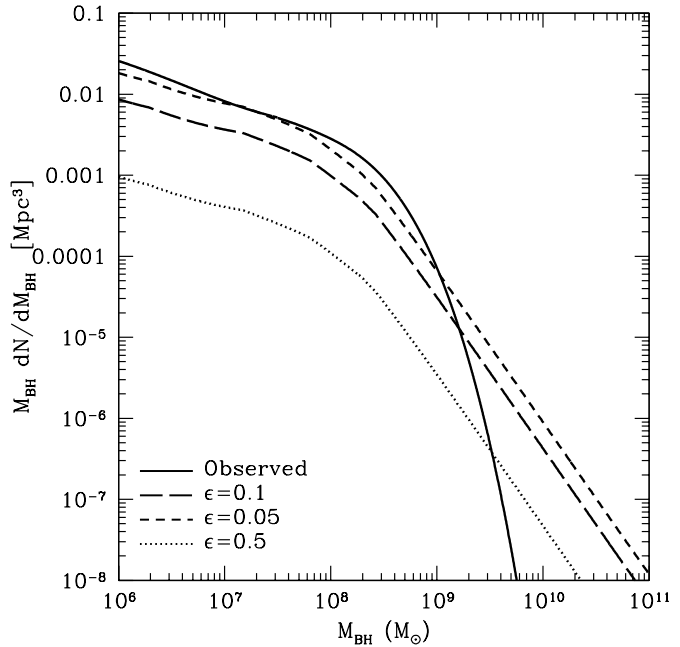}
\caption{Black hole spatial density per unit mass as function of black
hole mass. Dashed lines show the values inferred by integrating the
hard X-ray Luminosity function of Ueda et al. (2003) using the
bolometric corrections described on the text for three different
efficiencies: 0.05 ({\it dashed}), 0.1 ({\it long dashed}) and 0.5 ({\it
dotted}). The {\it solid line} shows the derived number density of BHs
from the SDSS local measured velocity function obtained using the
Merritt \& Ferrarese correlation between the black hole mass and
velocity dispersion of bulges.  }
\end{center}
\end{figure}

As can be seen in Fig.~2, for a reasonable value of the efficiency,
$\epsilon\,\gtrsim\,0.05$, there is a good agreement between the BH
mass density at $z=0$, as obtained from the velocity dispersion of
bulges, and the density inferred from AGN relics, for BH masses
smaller than $\sim 2-3\,\times\,10^{ 9}\,M_{\sun}$. However, for higher
masses, in particular the UMBH mass range, independent of the value of
$\epsilon$ assumed, the BH mass density from AGN relics is
significantly higher than the observed value, indicating that UMBHs
should be more abundant than current observations suggest. If there is
a mass dependent efficiency factor for accretion such that higher mass
BHs tend to accrete at higher efficiency and hence at lower rates,
then our estimate of the high mass tail would be an
over-estimate. There is however no evidence for such a mass dependence
at lower masses (Hopkins, Narayan \& Hernquist 2006).

The SDSS First Data Release covers approximately 2000 square degrees
(Abazajian et al. 2003), yielding a comoving volume of a cone on the
sky out to $z = 0.3$ of $3.34 \times 10^8$ Mpc$^3$. Given our
predicted abundance above, we expect $\sim$ 1000 UMBHs in the SDSS
volume, however only a few are detected. {\bf No combination of assumed
accretion efficiency and Eddington ratio coupled with the X-ray AGN LF
can reproduce the observed local abundance at the high mass end.}

\subsection{Evidence for an upper limit to black hole masses}

However, we find that modifying one of the key assumptions made above
brings the predicted abundance of local UMBHs into better agreement
with current observations. In the modeling we have extrapolated the
observed X-ray AGN LF slope to brighter luminosities. We find that if
this slope is steepened at the bright end, we can reproduce the
observed UMBH mass function at $z = 0$ for $M\geq$10$^9M_\odot$ as
well.  In order to reconcile the observationally derived local black
hole mass function at the high mass end, the slope $\gamma_2$ in
eqn. (1) needs to be modified. We find that the slope $\gamma_2$ for
black hole masses $M_{\rm bh} < 10^9\,\msun$ is $\sim\,2.2$, which
however, does not provide a good-fit for higher masses. A slope
steeper than $\gamma_2 = 5$ is required to fit BH masses in excess of
$10^9$, we find that formally the best-fit is found in
reduced-$\chi^2$ terms for the value of $\gamma_2 = 6.9$. {\bf Such a
steepening simulates the cut-off of a self-regulation mechanism that
limits black hole masses and sets in at every epoch}.  

In Fig.~3, the results of such a self-limiting growth model are
plotted. The predicted abundance of UMBHs is now in much better
agreement with observations at $z = 0$ and is consistent with the
number of UMBHs detected by SDSS. In a self-regulated mass growth
model we predict the abundance of UMBHs at $z = 0$ to be $7 \times
10^{-7}\, {\rm Mpc^{-3}}$.  Therefore, requiring consistency between
the X-ray and optical views of black hole growth and assembly with the
observed number of UMBHs at $z = 0$, points to the existence of a
self-regulation mechanism that limits BH masses. The self-regulation
is implemented as a steepening of the X-ray AGN LF at the luminous end
(the value of $\gamma_2$ needed is plotted in Fig.~4) and does not
have an important effect on the X-ray background, since this change in
slope only affects sources with X-ray luminosities $L(2-10\,keV)$
greater than 10$^{45}$~erg~s$^{-1}$, while most of the X-ray
background emission is produced by sources with luminosities of
10$^{43-44}$~erg~s$^{-1}$, as found by Treister \& Urry (2005). Note
that the use of a bolometric luminosity function, as reported by
Hopkins, Richards \& Hernquist (2007) does not match the slope at the
high mass end (shown as the dashed curve in Fig.~3). One key
consequence of our model is that the slope of the $M_{\rm bh} -
\sigma$ relation at the high mass end likely evolves with
redshift. Recent cosmological simulations find evidence for such a
trend (Di Matteo et al 2008). The functional form of this expected
variation will depend on the specific model of self-regulation
employed. Below, we explore physical processes that are likely to
regulate the growth of BHs in galactic nuclei.

\begin{figure}
\begin{center}
\includegraphics[height=9cm,width=9cm]{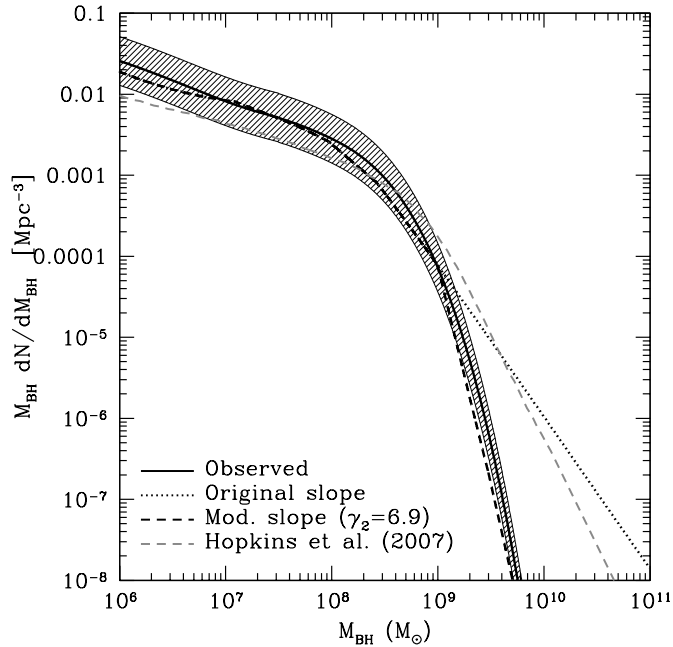}
\caption{Black hole spatial density per unit mass as function of black
hole mass. The {\it solid line} shows the SDSS-derived values, as
shown in Fig.~2, assuming a constant 30\% uncertainty ({\it shaded
region}). The {\it dotted line} shows the values derived integrating
the hard X-ray luminosity function for an efficiency of 0.05, while
the {\it gray dashed line} shows the relation reported by Hopkins,
Richards \& Hernquist (2007; fig 10) using a bolometric luminosity
function. In order to match the observed relation, the slope of the
hard X-ray luminosity function was modified for masses higher than
$10^9$M$_\odot$, as shown by the {\it black dashed line}.}
\end{center}
\end{figure}

Converting the high end of the local black hole mass function into the
equivalent velocity dispersions of the host spheroids we find values
in excess of 350 kms$^{-1}$.  In the context of the currently popular
hierarchical model for the assembly of structure, the most massive
galaxies in the Universe are expected to be the central galaxies in
clusters. High-$\sigma$ peaks in the density fluctuation field at
early times seed clusters that assemble at later times, and hence
these are the preferred locations for the formation of the most
massive galaxies in a cold dark matter dominated Universe.

\begin{figure}
\begin{center}
\includegraphics[height=9cm,width=9cm]{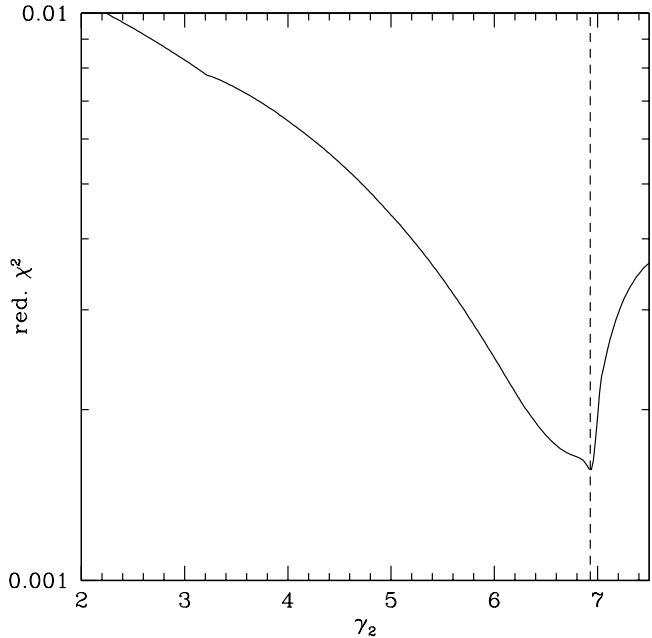}
\caption{The reduced $chi^2$ for the index $\gamma_2$ in the X-ray AGN
  LF required to match the high mass end of the local black hole mass
  density. In order to match the observed relation, the slope of the
hard X-ray luminosity function was modified for masses higher than
$10^9$M$_\odot$.}
\end{center}
\end{figure}

\section{The upper limit to BH masses from self-regulation arguments}

While we predict above that a few, rare UMBHs are likely to exist at
the centers of the brightest central galaxies in clusters, we further
argue that there likely exists an upper limit to black hole
masses. Evidence for this is presented using several plausible
physical scenarios that attempt to explain the coeval formation of the
black hole and the stellar component in galactic nuclei. Clearly the
existence of UMBHs is intricately related to the highest mass galaxies
that can form in the Universe.

Given that star formation and black hole fueling appear to be coupled
(e.g. di Matteo et al. 2005 and references therein; Silk \& Rees
1998), it is likely that there is a self-limiting growth cycle for BHs
and therefore a physical upper limit to their masses. Here we present
several distinct arguments that can be used to estimate the final
masses of BHs (Haehnelt, Natarajan \& Rees 1998; Silk \& Rees 1998;
Murray, Quataert \& Thompson 2004 and King 2005). These involve
self-limiting growth due to a momentum-driven wind, self-limiting
growth due to the radiation pressure of a momentum-driven wind, and
from an energy-driven superwind model .

Murray, Quataert \& Thompson (2004) argue that the feedback from
momentum driven winds, limits the stellar luminosity, which in turn
regulates the BH mass. They argue for Eddington limited star formation
with a maximum stellar luminosity,
\begin{eqnarray}
L_M = \frac{4 f_g c}{G}\,\sigma^4,
\end{eqnarray}
where, $f_g$ is the gas fraction in the halo and $\sigma$ the velocity
dispersion of the host galaxy. Star formation in this scheme is
unlikely to evacuate the gas at small radius in the galactic nucleus,
therefore, all the gas in the inner-most regions fuel the BH. The
growing BH itself clears out this nuclear region with its accretion
luminosity approaches $L_M$. At this point the fuel supply to the BH
is shut-off and this may shut off the star formation as well. The
final BH mass is then given by,
\begin{eqnarray}
M_{\rm BH} = \frac{f_g \kappa_{\rm es}}{\pi G^2}\,\sigma^4,
\end{eqnarray}
where $\kappa_{\rm es}$ is the electron scattering opacity.  For the
most massive, nearby early-type galaxies at the very tail of the
measured SDSS velocity dispersion function with velocity dispersions
of $\sim$ 350 - 400 kms$^{-1}$ (Bernardi et al. 2005) this gives a
final BH mass of $\sim\,10^{10} \msun$. Therefore, normal galaxies
with large velocity dispersions are the presumptive hosts for
UMBHs.\footnote{Objects with high velocity dispersion as a consequence
  of superposition are not the hosts of UMBHs} Furthermore, there
appears to be a strong indication of the existence of an upper mass
limit for accreting black holes derived from SDSS DR3 by Vestergard et
al. (2008) in every redshift bin from $z = 0.3 - 5$.

An alternative upper limit can be obtained when the emitted energy
from the accreting BH back reacts with the accretion flow itself
(Haehnelt, Natarajan \& Rees 1998).  The final shut-down of accretion
will depend on whether the emitted energy can back-react on the
accretion flow prior to fuel exhaustion. This arguement provides a limit,
\begin{eqnarray}
M_{\rm bh} &\sim& 5.6 \times 10^9\msun\,  (f_{\rm kin}/0.0001)^{-1}\,
 j_{\rm d}^{-5} \,   
\left( \frac {\lambda}{0.05} \right )^{-5} \, 
\left (\frac {m_{\rm d}}{0.1} \right )^{5} \, \nonumber\\ 
&&\qquad \qquad \times
\left (\frac {\sigma}{350\, \kms}\right )^{5} \msun,
\end{eqnarray}
where $f_{\rm kin}$ is the fraction of the accretion luminosity which
is deposited as kinetic energy into the accretion flow (cf. Silk \&
Rees 1998), $\lambda$ is the spin parameter of the DM halo, $j_d$ is
the specific angular momentum of the disk, $m_d$ is the disk mass
fraction. The back-reaction timescale will be related to the dynamical
timescale of the outer parts of the disk and/or the core of the DM
halo and should set the duration of the optically bright phase.  It is
interesting to note here that the accretion rate will change from
super-Eddington to sub-Eddington without much gain in mass if the
back-reaction timescale is shorter than the Salpeter time. The overall
emission efficiency is then determined by the value of $\dot m$ when
the back-reaction sets in and is reduced by a factor $1/\dot m$
compared to accretion at below the Eddington rate. By substituting the
value of the velocity dispersion of nearby cD's $\sim 350\,{\rm
kms^{-1}}$, we obtain a limiting value of the mass, if we assume that
the bulk of the mass growth occurs in the optically bright quasar
phase. Due to the dependence on the spin parameter $\lambda$ of the DM
halo, the desired UMBH mass range can arise preferentially in high
velocity dispersion halos with low spin. \footnote{The distribution of
spins of DM halos measured from N-body simulations is found to be a
log-normal with a median value of 0.05, and since there is no
significant halo mass dependence, a small fraction of the halos do
reside in this low-spin tail.}

King (2005) presents a model that exploits the observed AGN-starburst
connection to couple black hole growth and star formation. As the
black hole grows, an outflow drives a shell into the surrounding gas
which stalls after a dynamical time-scale at a radius determined by
the BH mass. The gas trapped inside this bubble cools, forms stars and
is recycled as accretion and outflow. Once the BH reaches a critical
mass, this region attains a size such that the gas can no longer cool
efficiently. The resulting energy-driven flow expels the remaining gas
as a superwind, thereby fixing the observed $M_{\rm bh} - \sigma$
relation as well as the total stellar mass of the bulge at values in
good agreement with current observations. The limiting BH mass is
given by:
\begin{eqnarray}
M_{\rm bh} = \frac{f_g\,\kappa}{\pi\,G^2}\,\sigma^4,
\end{eqnarray}
where $f_g$ is the gas fraction ($\Omega_{baryon}/\Omega_{matter} =
0.16$, $\kappa$ the electron scattering opacity and $\sigma$ the
velocity dispersion. This model argues that black hole growth
inevitably produces starburts and ultimately a superwind.

Note that both the Murray, Quataert \& Thompson (2004) model and the
King (2005) model predict $M_{\rm bh} \propto \sigma^{4}$ while the
Haehnelt et al. (1998) and Silk \& Rees (1998) predict a $\sigma^5$
dependence. The current error bars on the observational mass estimates
for black holes preclude discrimination between these two
possibilities. Shutdown of star formation above a critical halo mass effected
by the growing AGN has also been proposed as a self-limiting mechanism
to cap BH growth and simultaneously explain the dichotomy in galaxy properties
(Croton et al. 2006; Cattaneo et al. 2006)

\section{Prospects for detection of quiescent UMBHs}

UMBHs are expected to be rare in the local Universe, from our analysis
of the X-ray luminosity function of AGN, we predict an abundance
ranging from $\sim$ few times $10^{-6} -- 10^{-7}\,{\rm
  Mpc}^{-3}$. These estimates are in good agreement with those
obtained from optical quasars in the SDSS DR3 by Vestergard et
al. (2008).  The results of the first attempts to detect and measure
masses for UMBHs is promising.  Dalla Bonta et al.  (2007) selected 3
Brightest Cluster Galaxies (BCGs) in Abell 1836, Abell 2052 and Abell
3565. Using ACS (Advanced Camera for Surveys) aboard the Hubble Space
Telescope and the Imaging Spectrograph (STIS), they obtained high
resolution spectroscopy of the $H\alpha$ and $N$II emission lines to
measure the kinematics of the central ionized gas. They present BH
mass estimates for 2 of these BCGs, $M_{\rm bh} = 4.8^{+0.8}_{-0.7}
\times 10^9\,M_{\odot}$ and $M_{\rm bh} = 1.3^{+0.3}_{-0.4} \times
10^9\,M_{\odot}$ and an upper limit for the BH mass on the third
candidate of $M_{\rm bh} \leq 7.3\times 10^{10}\,M_{\odot}$.

It is interesting to note that Bernardi et al. (2005) in a census of
the most massive galaxies in the SDSS survey do find candidates with
large velocity dispersions $(\geq 350\,{\rm kms^{-1}})$. The largest
systems they find are claimed to be extremes of the early-type galaxy
population, as they have the largest velocity dispersions.  These
$\sim 31$ systems (see Table 1 of Bernardi et al. (2006) for details
on these candidates) are not distant outliers from the Fundamental
Plane and the mass-to-light scaling relations defined by the bulk of
the early-type galaxy population. Clear outliers from these scaling
relations tend to be objects in superposition for which they have
evidence from spectra and images. We argue that these extreme
early-type galaxies might harbour UMBHs and likely their abundance
offers key constraints on the physics of galaxy formation. Although
the observations are challenging, a more comprehensive and systematic
survey of nearby BCGs is likely to yield our first local UMBH before
long. As discussed above, candidates from the SDSS are promising
targets for observational follow-up as they are extremely
luminous. Utilizing the Hubble Space Telescope, the light profile
might show evidence for the existence of an UMBH in the center
(e.g. Lauer et al. 2002). In fact, for $SDSS\,J032834.7+001050.1$ and
$SDSS\,J161541.3+471004.3$, it may be possible to measure spatially
resolved velocity dispersion profiles even from ground-based
facilities.

\section{Discussion}

The interplay between the evolution of BHs and the hierarchical
build-up of galaxies appears as scaling relations between the masses
of BHs and global properties of their hosts such as the BH mass
vs. bulge velocity dispersion - the $M_{\rm bh} - \sigma_{\rm bulge}$
relation and the BH mass vs. bulge luminosity $M_{\rm bh} - L_{\rm
Bulge}$ relation. The low BH mass end of this relation has recently
been probed by Ferrarese et al. (2006) in an ACS survey of the Virgo
cluster galaxies. They find that galaxies brighter than $M_B \sim -20$
host a supermassive central BH whereas fainter galaxies host a central
nucleus, referred to as a central massive object (CMO). Ferrarese et
al. report that a common $M_{\rm CMO} - M_{\rm gal}$ relation leads
smoothly down from the scaling relations observed for more more
massive galaxies. Extrapolating observed scaling relations to higher
BH masses to the UMBH range, we predict that these are likely hosted
by the massive, high luminosity, central galaxies in clusters with
large velocity dispersions. The velocity dispersion function of
early-type galaxies measured from the SDSS points to the existence of
a high velocity dispersion tail with $\sigma > 350\,{\rm kms}^{-1}$
(Bernardi et al. 2006). If the observed scaling relations extend to
the higher mass end as well, these early-types are the most likely
hosts for UMBHs.

Recent simulation work that follows the merger history of cluster
scale dark matter halos and the growth of BHs hosted in them by Yoo et
al.(2007) also predict the existence of a rare population of local
UMBHs.  However, theoretical arguments suggest that there may be an
upper limit to the mass of a BH that can grow in a given galactic
nucleus hosted in a dark matter halo of a given spin. Clearly the
issue of the existence of UMBHs is intimately linked to the efficiency
of galaxy formation and the formation of the largest, most luminous
and massive galaxies in the Universe.

Possible explanations for the tight correlation observed between the
velocity dispersion of the spheroid and black hole mass involve a
range of self-regulated feedback prescriptions. An estimate of the
upper limits on the black hole mass that can assemble in the most
massive spheroids can be derived for all these models and they all
point to the existence of UMBHs.

In this paper, we have argued that while rare UMBHs likely exist,
there is nevertheless an upper limit of $\sim 10^{10}\,\msun$ for the
mass of BHs that inhabit galactic nuclei in the Universe. We first
show that our current understanding of the accretion history and mass
build up of black holes allows and implies the existence of UMBHs
locally. This is primarily driven by new work that predicts the
formation of massive black hole seeds at high redshift (Lodato \&
Natarajan 2007) and their subsequent evolution (Volonteri, Lodato \&
Natarajan 2008). Starting with massive seeds and following their
build-up through hierarchical merging in the context of structure
formation in a cold dark matter dominated Universe, we show that a
viable pathway to the formation of UMBHs exists. There is also
compelling evidence from the observed evolution of X-ray AGN for the
existence of a local UMBH population. Convolving the observed X-ray
LF's of AGN, with a simple accretion model, the mass function of black
holes at $z = 0$ is estimated. Mimic-ing the effect of self-regulation
processes that impose an upper limit to BH masses and incorporating
this into the X-ray AGN LF we find that the observed UMBH mass
function at $z = 0$ is reproduced. This self-regulation limited growth
is implemented by steepening the high luminosity end of the AGN LF at
the bright end. We estimate the abundance of UMBHs to be $\sim \,7
\times 10^{-7}\, Mpc^{-3}$ at $z = 0$. The key prediction of our model
is that the slope of the $M_{\rm bh} - \sigma$ relation likely evolves with
redshift at the high mass end. Probing this is observationally challenging at
the present time but there are several bright, massive early-type
galaxies that are promising host candidates from the SDSS survey as
well as a survey of bright central galaxies of nearby
clusters. Observational detection of UMBHs will provide key insights
into the physics of galaxy formation and black hole assembly in the
Universe.

\section*{Acknowledgments} 

We thank Steinn Sigurdsson and Meg Urry for useful discussions.

\end{document}